\documentclass[11pt]{article}

\usepackage[margin=1in]{geometry}

\usepackage[utf8]{inputenc}
\usepackage[T1]{fontenc}
\usepackage{hyperref}
\usepackage{url}
\usepackage{booktabs}
\usepackage{amsfonts}
\usepackage{amsmath}
\usepackage{nicefrac}
\usepackage{microtype}
\usepackage{xcolor}
\usepackage{graphicx}
\usepackage{multirow}
\usepackage{array}
\usepackage{algorithm}
\usepackage{algorithmic}
\usepackage{natbib}

\usepackage{enumitem}
\usepackage{setspace}

\title{Cogniscope: A Synthetic Longitudinal Benchmark and
Browser-Based Evaluation Framework for Early-Risk
Cognitive AI Systems}

\author{
  Mahfuza Farooque\thanks{Corresponding author: \texttt{mff5187@psu.edu}} \\
  Pennsylvania State University
  \and
  Ananya Drishti \\
  Pennsylvania State University
  \and
  Uttkarsh Agarwal \\
  Pennsylvania State University
  \and
  Zahra Abdul Basit \\
  Pennsylvania State University
  \and
  Asish Kondragunta \\
  Pennsylvania State University
  \and
  Mukhil Muruganantham Prakaash  \\
  Pennsylvania State University
}

\date{}

\begin{document}
\maketitle

\begin{abstract}
We present \textbf{Cogniscope}, an open evaluation framework for
studying longitudinal early-risk AI systems under controlled
behavioral drift, sparse observations, delayed evidence, and
heterogeneous progression patterns. Cogniscope combines two
complementary components: a synthetic simulation engine that
generates privacy-preserving longitudinal behavioral traces
aligned with configurable latent risk trajectories, and a
browser-based data-collection instrument implemented as a Chrome
extension for capturing naturalistic video interaction telemetry
and micro-question responses during YouTube playback. The released
benchmark includes 200,000 simulated video-interaction records
from 200 users over 200 days, a 504-session schema-aligned
synthetic deployment dataset across nine behavioral profiles, an
18-table relational schema, baseline evaluation scripts, and
time-aware metrics including Early Risk Detection Error (ERDE)
and time-to-detection (TTD). We emphasize that Cogniscope is
\emph{not} a diagnostic system and does not claim clinical
validity. Instead, it provides a reusable testbed for evaluating
how sequential models behave under known longitudinal challenges
before deployment with real human-subject data. Experiments show
that simple behavioral coherence signals separate simulated risk
states under controlled priors, while rule-based
deployment-profile classification remains challenging, motivating
learned temporal models and robust evaluation protocols. 
\end{abstract}

\section{Introduction}
\label{sec:intro}

Alzheimer's disease (AD) is the leading cause of dementia
worldwide, characterized by progressive decline in memory,
language, and attentional control~\citep{jack2018niaa,
DeLeon2007AnnNYAS}. Its prodromal stage, Mild Cognitive
Impairment (MCI), represents a critical window in which cognitive
changes are detectable while functional independence is largely
preserved~\citep{petersen2014lancet}. Despite this, 75\% of
individuals with early cognitive decline go
undiagnosed~\citep{gauthier2021}. Conventional approaches, such
as MRI, cerebrospinal fluid analysis, and structured
neuropsychological batteries, are expensive, episodic, and
inaccessible at scale~\citep{Mueller2005ADNI,yang2022}.

Two complementary directions have emerged. First, \emph{digital
phenotyping} has demonstrated that speech-based markers carry
detectable early signals of cognitive
decline~\citep{Fraser2015JAD,Karlekar2018Arxiv,Mueller2018JCEN,
Cummings2019PragSoc,Balagopalan2021Frontiers,Agbavor2022PLoS},
with state-of-the-art classifiers achieving $F_1 = 0.873$ on
dedicated elicitation datasets~\citep{pahar2025cognospeak}.
Second, \emph{passive interaction telemetry}, including pauses,
rewinds, skips, and comprehension-question responses during
everyday media consumption, offers an ecologically valid,
low-burden complement~\citep{Seelye2015DADM,Dodge2015TRCI,
robin2021}.

Despite this progress, the evaluation ecosystem remains
fragmented. Existing benchmarks
(ADReSS~\citep{luz2020adress}, ADReSSo~\citep{luz2021adresso},
DementiaBank~\citep{becker1994}) require dedicated elicitation
sessions and lack reusable executable artifacts, documented
feature derivation pipelines, and population-specific evaluation
protocols. No existing resource simultaneously provides a
configurable simulation engine, a deployable data-collection
instrument, a structured longitudinal schema, and a synthetic
benchmark dataset.

\textbf{Cogniscope} addresses all of these gaps through an
integrated two-level architecture: a \emph{simulation level} for
privacy-preserving controlled experimentation, and a
\emph{deployment level} realized as a Chrome extension backed by
a production-grade relational schema.

\paragraph{Contributions.}
\begin{enumerate}[noitemsep,topsep=2pt,leftmargin=*]
  \item A \textbf{configurable simulation and benchmarking
    framework} generating longitudinal interaction traces aligned
    with latent risk trajectories under drift, sparsity, and
    heterogeneous progression.
  \item A \textbf{time-aware evaluation protocol} using ERDE,
    TTD, and fixed false-positive operating points.
  \item A \textbf{publicly available Chrome extension
    implementation} for capturing video interaction telemetry and
    LLM-generated comprehension questions during YouTube playback.
  \item An \textbf{18-table relational schema}
    (Appendix~\ref{app:schema_full}) supporting longitudinal
    monitoring, population-specific pathways, ground-truth
    labeling, and responsible review workflows.
  \item A \textbf{synthetic evaluation dataset} of 504 simulated
    sessions across nine behavioral profiles with transparent
    rule-based labels (Appendix~\ref{app:classifier}).
  \item A \textbf{responsible design specification} separating
    neutral user-facing statistics from restricted reviewer-only
    prediction outputs.
\end{enumerate}

\paragraph{Scope of claims.}
Cogniscope is intended as a controlled evaluation environment,
not as a clinically validated cognitive assessment tool. The
released datasets are synthetic and are designed to support
stress-testing, debugging, benchmarking, and reproducibility
studies for longitudinal early-risk AI systems. The clinical
terminology used in this paper provides motivation and task
structure, but the labels should be interpreted as
\emph{simulated risk states} rather than diagnostic categories.
Accordingly, our evaluation claims concern whether models can
operate under controlled temporal challenges such as drift,
sparse evidence, delayed observations, and heterogeneous
trajectories. We do not claim that the reported models detect
Alzheimer's disease, mild cognitive impairment, or any clinical
condition in real users.

\paragraph{Evaluations \& Datasets track fit.}
Cogniscope is designed for the Evaluations and Datasets setting:
the primary contribution is not a new predictive model, but a
reusable evaluation substrate. It enables controlled comparison
of longitudinal models under predefined sources of uncertainty,
provides transparent labels and generation rules for
auditability, and supplies a deployable collection instrument so
that future real-world datasets can follow the same schema and
feature vocabulary. While AD serves as the motivating case study,
the framework generalizes to any early-risk monitoring technology
relying on longitudinal behavioral and language signals.

\section{Related Work}
\label{sec:related}

\paragraph{Digital biomarkers and existing benchmarks.}
Speech-based biomarkers, including linguistic coherence,
disfluency, lexical diversity, and semantic drift, have
demonstrated utility in distinguishing MCI and
AD~\citep{Cummings2019PragSoc,Karlekar2018Arxiv,Agbavor2022PLoS,
Fraser2015JAD,Balagopalan2021Frontiers,pahar2025cognospeak}.
Complementary work explores cursor dynamics and device usage as
passive markers of functional
change~\citep{Seelye2015DADM,Dodge2015TRCI}.
ADReSS~\citep{luz2020adress}, ADReSSo~\citep{luz2021adresso},
DementiaBank~\citep{becker1994,Lanzi2023}, and
eRisk~\citep{LosadaCrestani2016} provide valuable benchmarks but
require dedicated elicitation sessions, lack simulation
environments, and do not capture naturalistic video interaction
behavior. Cogniscope complements rather than replaces these
datasets.

\paragraph{Passive sensing and simulation-based evaluation.}
Motor-spatial telemetry and social media engagement signals have
been proposed as ecologically grounded proxies for attentional
load and working-memory
demands~\citep{yang2022,globalmediainsight2025,umbrex2025,
Milne2022BigDataSoc,Onnela2025,Ali2024Cureus,robin2021}.
Simulation-based evaluation offers a principled means of probing
system behavior under controlled perturbations prior to
real-world
deployment~\citep{Vlontzou2025SciRep,Patil2025MethodsX,
Lee2025SciRep,saltelli2019,Topol2019NatMed,javed2024air}.
LLM-based MCQ generation from transcripts has been applied in
educational settings~\citep{openai2023gpt4}; Cogniscope
integrates this directly into the assessment pipeline. Ethical
analyses stress that early predictive systems must carefully
manage uncertainty and avoid overinterpretation of preclinical
signals~\citep{Smedinga2018JAD,blasi2005assessment}, a
principle enforced throughout Cogniscope's design.
Unlike ADReSS, ADReSSo, DementiaBank, and eRisk, Cogniscope is
not positioned as a clinically validated human-subject dataset;
its contribution is an executable synthetic evaluation
environment paired with a deployable collection instrument,
schema, and time-aware benchmark protocol.
Appendix~\ref{app:prior_work} summarizes representative prior
resources and Cogniscope's relationship to them.

\section{Framework Architecture}
\label{sec:architecture}

\begin{figure*}[t]
  \centering
  \includegraphics[width=0.9\linewidth]{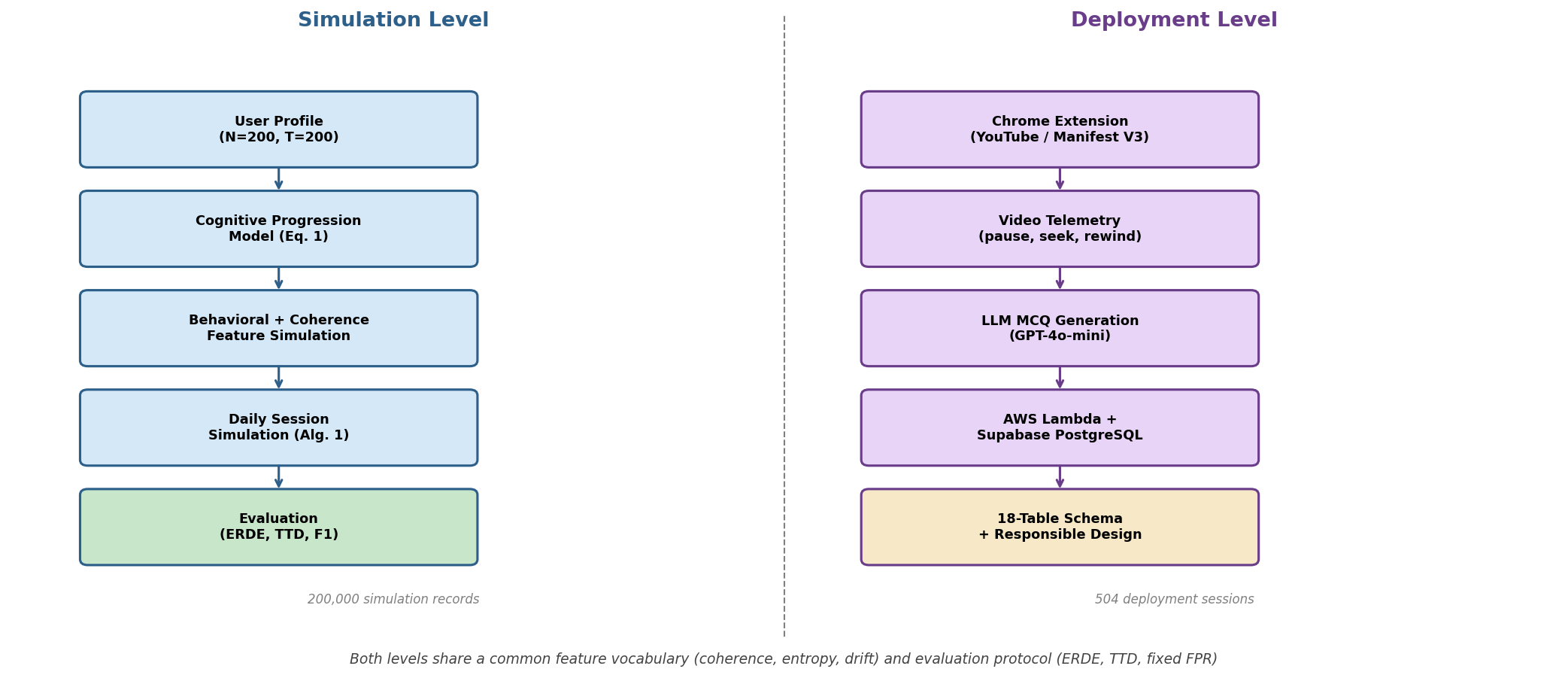}
  \caption{Cogniscope two-level architecture. Simulation level
    (left) generates synthetic behavioral traces for controlled
    benchmarking. Deployment level (right) captures interaction
    telemetry via Chrome extension for future real-world data
    collection. Both levels share a common feature vocabulary
    and evaluation protocol.}
  \label{fig:architecture}
\end{figure*}

Cogniscope operates at two complementary levels
(Figure~\ref{fig:architecture}). The \emph{simulation level}
(Section~\ref{sec:simulation}) provides a configurable engine
for generating privacy-preserving longitudinal behavioral traces
under controlled conditions. The \emph{deployment level}
(Section~\ref{sec:deployment}) provides an instrument for
capturing naturalistic interaction data in future real-world
studies. Both levels share a common feature vocabulary and
evaluation protocol (Section~\ref{sec:features}), enabling
comparison between simulated traces and future real-world data
collected under the same schema.

\section{Simulation Level}
\label{sec:simulation}

\subsection{Synthetic Data Generation}

We simulate a cohort of $N=200$ users over $T=200$ days. Each
user participates in one daily session of short-form video
consumption (five videos, 15--90 seconds, from five topical
categories) followed by lightweight language-based probes:
(i)~a free-form summary (1--3 sentences) and (ii)~short
comprehension or affective responses. Textual responses are
generated via a Groq API (Llama-3-based model) with
template-based fallback. Cognitive state labels parameterize
latent feature distributions only; all reported results use a
\emph{no-label generation mode} in which prompts do not expose
cognitive state information.

\subsection{Cognitive Progression Modeling}

We use labels such as Healthy, MCI, and EarlyAD as shorthand
for \emph{simulated latent risk states} inspired by
cognitive-monitoring scenarios. These labels do not correspond
to clinically adjudicated diagnoses and should not be
interpreted as medical ground truth. Each user $u$ is assigned
a latent progression profile $P_u$ governing transitions across
five ordered risk states via ordered transition points
$(D_3, D_4, D_5, D_6)$:
\begin{equation}
L_u(d) =
\begin{cases}
\text{Healthy}, & d < D_3,\\
\text{MCI},     & D_3 \le d < D_4,\\
\text{EarlyAD}, & D_4 \le d < D_5,\\
\text{ModAD},   & D_5 \le d < D_6,\\
\text{SevAD},   & d \ge D_6.
\end{cases}
\label{eq:progression}
\end{equation}
Transition points are independently sampled per user, inducing
heterogeneous trajectories capturing variability in onset timing
and decline velocity.

\subsection{Behavioral and Linguistic Feature Simulation}

\paragraph{Behavioral features.}
Engagement is modeled through state-dependent distributions over
watch time, skip duration, pausing, replay behavior, reaction
latency, and interaction frequency. Table~\ref{tab:behaviors}
lists ranges per cognitive state; features are sampled
independently per interaction and logged at the video level.

\begin{table*}[t]
\centering
\footnotesize
\setlength{\tabcolsep}{3.5pt}
\caption{Behavioral metric distributions by simulated risk state
  (\textit{per-session probabilities}). Priors are illustrative;
  not direct estimates of platform logs.}
\label{tab:behaviors}
\begin{tabular}{lccccccccr}
\toprule
\textbf{Label} &
\textbf{WT (s)} &
\textbf{Skip (s)} &
\textbf{Pause} &
\textbf{Replay} &
\textbf{RT (s)} &
\textbf{Like (\%)} &
\textbf{Share (\%)} &
\textbf{Churn (\%)} &
\textbf{Logins/d} \\
\midrule
Healthy & 60--75 & 0--5   & 0--2 & 0--1 & 4--6
        & 30--40 & 15--20 & 1    & 2--3 \\
MCI     & 40--60 & 5--15  & 1--3 & 1--3 & 7--10
        & 20--25 & 10--15 & 2--3 & 1--2 \\
EarlyAD & 20--40 & 10--25 & 2--5 & 2--5 & 11--14
        & 5--10  & 3--7   & 5--6 & 0.5--1 \\
ModAD   & 15--25 & 15--30 & 3--6 & 3--6 & 14--17
        & 2--5   & 1--3   & 7--8 & 0.3--0.8 \\
SevAD   & 10--15 & 20--40 & 4--8 & 4--8 & 18--22
        & 0--2   & 0--1   & 12--15 & $<$0.5 \\
\bottomrule
\end{tabular}
\end{table*}

\paragraph{Coherence features.}
In the absence of raw textual summaries, semantic coherence is
approximated using response-level behavioral signals, which have
been shown to correlate with cognitive processing
stability~\citep{Seelye2015DADM,robin2021}. Coherence is
computed as a weighted combination of accuracy, response latency
efficiency, skip rate, and response consistency:
\begin{equation}
C_{u,d} = w_1 \cdot \mathrm{acc}_{u,d}
        + w_2 \cdot e^{-t_{u,d}/60}
        + w_3 \cdot (1 - \mathrm{skip}_{u,d})
        + w_4 \cdot \mathrm{cons}_{u,d},
\label{eq:coherence}
\end{equation}
where $w_1{=}0.4$, $w_2{=}0.2$, $w_3{=}0.2$, $w_4{=}0.2$.
Global semantic drift is defined as $\Delta C_{u,d} = 1 -
C_{u,d}$. Gaussian noise ($\sigma \in \{0.05, 0.1, 0.2,
0.3\}$) is injected into coherence components to model
day-to-day variability. User sessions are generated following
Algorithm~\ref{alg:simulation} (Appendix~\ref{app:algorithms});
latent states are withheld from downstream models and retained
only for evaluation.

\section{Deployment Level}
\label{sec:deployment}

\subsection{Chrome Extension and Backend}

\begin{figure}[t]
  \centering
  \includegraphics[width=0.56\linewidth]{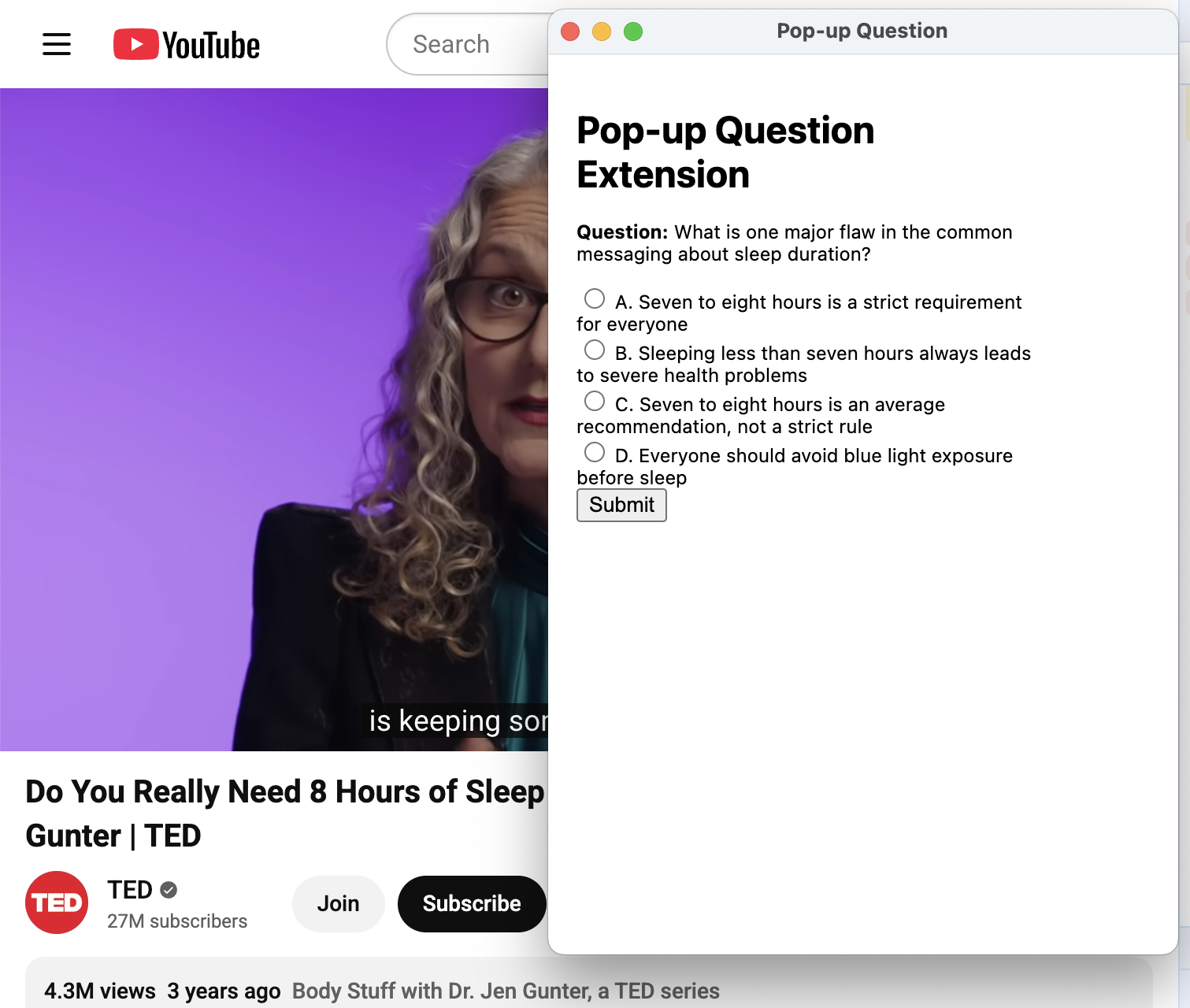}
  \caption{Cogniscope Chrome extension: pop-up comprehension
    question during YouTube playback. Designed for future
    real-world data collection; all datasets in this paper
    are synthetic.}
  \label{fig:popUp}
\end{figure}

Cogniscope is implemented as a Manifest V3 Chrome extension. A
content script monitors YouTube DOM events and the player API to
capture playback events (\texttt{pause}, \texttt{play},
\texttt{seek}, \texttt{rewind}, \texttt{ended}) with timestamps,
session metadata (video ID, title, duration, device type), and
micro-question responses (popup MCQs at transcript-derived
checkpoints with answer selection, correctness, and response
latency). Figure~\ref{fig:popUp} shows the extension interface.
The backend is a serverless AWS architecture (API Gateway +
Lambda) persisting data in a Supabase-hosted PostgreSQL
database. Sensitive prediction outputs are restricted via
row-level security policies to authorized reviewer roles. The
Chrome extension serves as a \emph{data-collection instrument}
for future real-world longitudinal studies; the evaluated
datasets in this paper are synthetic.

\subsection{LLM-Based Question Generation}

Transcript text is sent to GPT-4o-mini via a two-prompt
strategy: a \emph{classification prompt} identifies the
learning domain (comprehension, reasoning, factual recall,
inference), and a \emph{generation prompt} produces a
structured MCQ grounded in transcript evidence. Questions are
validated for unique options, a single correct answer, and
evidence traceability before storage.

\section{Relational Schema}
\label{sec:schema}

The Cogniscope schema comprises 18 tables organized into five
layers (Figure~\ref{fig:schema}); full DDL is in
Appendix~\ref{app:ddl} and the complete table inventory in
Appendix~\ref{app:schema_full}. The \textbf{User Layer}
(\texttt{user\_info}, \texttt{population\_profile}) assigns
users to monitoring pathways. The \textbf{Session Layer}
(\texttt{video\_info}, \texttt{video\_content\_metadata})
stores per-session watch records and content properties to
prevent confounding user difficulty with content difficulty.
The \textbf{Interaction Layer}
(\texttt{video\_interaction\_events}, \texttt{user\_response},
\texttt{question\_presentation\_log}) stores raw telemetry,
MCQ responses, and display context. The \textbf{Assessment
Layer} (\texttt{question\_bank}, \texttt{assessment\_schedule},
\texttt{assessment\_result}) supports structured baseline and
longitudinal follow-up. The \textbf{Modeling \& Safety Layer}
holds engineered features, rolling change metrics, interpretable
risk indicators, safety-action logs, and restricted prediction
outputs; \texttt{frontend\_summary\_view} exposes only neutral
statistics to end users.

\begin{figure}[t]
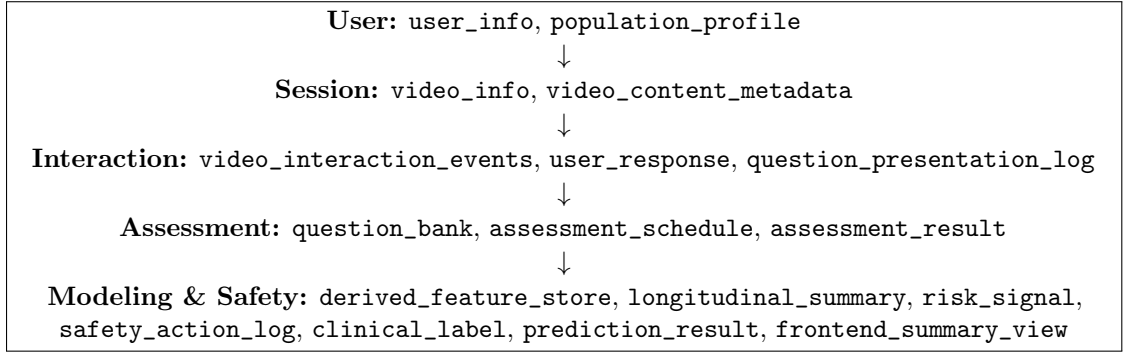

\centering
\fbox{%
\parbox{0.88\columnwidth}{%
\centering\small
\textbf{User:} \texttt{user\_info},
\texttt{population\_profile}\\[1pt]
$\downarrow$\\[1pt]
\textbf{Session:} \texttt{video\_info},
\texttt{video\_content\_metadata}\\[1pt]
$\downarrow$\\[1pt]
\textbf{Interaction:} \texttt{video\_interaction\_events},
\texttt{user\_response},
\texttt{question\_presentation\_log}\\[1pt]
$\downarrow$\\[1pt]
\textbf{Assessment:} \texttt{question\_bank},
\texttt{assessment\_schedule},
\texttt{assessment\_result}\\[1pt]
$\downarrow$\\[1pt]
\textbf{Modeling \& Safety:}
\texttt{derived\_feature\_store},
\texttt{longitudinal\_summary}, \texttt{risk\_signal},
\texttt{safety\_action\_log}, \texttt{clinical\_label},
\texttt{prediction\_result},
\texttt{frontend\_summary\_view}
}}
\caption{Cogniscope five-layer schema hierarchy (18 tables;
  full inventory in Appendix~\ref{app:schema_full}).}
\label{fig:schema}
\end{figure}

\section{Feature Derivation and Evaluation Protocol}
\label{sec:features}

\subsection{Shared Feature Vocabulary}

Both levels produce features from the same vocabulary. Each
user is represented as $\mathbf{X}_u =
\{x_{u,1},\ldots,x_{u,T}\}$, where $x_{u,t}$ jointly encodes:
\textbf{Coherence:} behavioral coherence score ($C_{u,d}$,
Eq.~\ref{eq:coherence}), semantic drift ($\Delta C_{u,d}$), and
component signals (accuracy, response latency, skip rate,
consistency). \textbf{Behavioral:} normalized watch time,
skip/pause behavior, replay frequency, response latency,
engagement counts; behavioral entropy $H_u = -\sum_i p_i \log
p_i$ summarizes interaction unpredictability.
\textbf{Assessment (deployment):} overall accuracy, retention
drop (immediate minus delayed accuracy), response-time
variability, video completion rate, missed-question rate, and
within-session accuracy consistency across early/middle/late
thirds.

\subsection{Temporal Risk Modeling}

Early-risk detection is formulated as a sequential prediction
problem: given prefix $x_{u,1:t}$, the model produces risk
score $r_{u,t} = f_\theta(x_{u,1:t})$. The released evaluation
scripts include GRU, TCN, lightweight Transformer, and
logistic-regression probes. In this paper, we report
coherence-threshold, ablation, and rule-based deployment-profile
results as reference baselines, with full temporal-model
benchmarking supported by the released code. All models use a
70/30 user-level split (no temporal leakage); hyperparameters
are in Appendix~\ref{app:hyperparams}.

\subsection{Time-Aware Evaluation Protocol}

Models are compared using: \textbf{ERDE} (Early Risk Detection
Error, penalizing late detection and false positives),
\textbf{TTD} (days from risk-state onset to first positive
detection), and \textbf{fixed FPR operating points} (1\%, 5\%,
10\%). Models observe only trajectory prefixes, reflecting
partial observability in real-world monitoring. The
deployment-level dataset uses nine transparent rule-based
learner-status categories (full taxonomy in
Appendix~\ref{app:taxonomy}).

\paragraph{Benchmark tasks.}
Cogniscope supports three primary benchmark tasks:
(i) prefix-based risk-state detection, where models predict
whether a user has entered a simulated risk state from
observations up to day $t$; (ii) early detection, where models
are evaluated by how soon they detect a transition after
simulated onset using ERDE and TTD; and (iii) robustness
evaluation, where models are tested under noise, missing
observations, delayed evidence, and held-out behavioral
profiles. These tasks evaluate longitudinal model behavior under
controlled uncertainty rather than clinical diagnostic accuracy.

\subsection{Recommended Challenge Splits}
\label{sec:splits}

In addition to the default user-level 70/30 split, Cogniscope
specifies four challenge splits, and the released data-generation
code allows users to instantiate them directly:
\textbf{(1)~Noise-shift split}, where models train on
low-noise trajectories and test on high-noise trajectories;
\textbf{(2)~Sparse-observation split}, where test users have
randomly missing daily sessions;
\textbf{(3)~Delayed-evidence split}, where labels are evaluated
only after a minimum observation window; and
\textbf{(4)~Profile-generalization split}, where models train
on a subset of behavioral profiles and test on held-out
profiles. These splits are intended to discourage overfitting
to simulation priors and to evaluate whether models remain
reliable under realistic longitudinal uncertainty.

\section{Datasets}
\label{sec:datasets}

\paragraph{Why synthetic data?}
Real longitudinal cognitive-risk datasets are difficult to
release because they involve sensitive health information, long
observation windows, and complex consent constraints. Cogniscope
uses synthetic data to make the evaluation environment fully
reproducible, privacy-preserving, and controllable. This allows
researchers to test model behavior under precisely specified
conditions, including transition timing, observation sparsity,
noise, drift, and profile heterogeneity. The synthetic
benchmark is therefore not a substitute for clinical
validation, but a \emph{pre-deployment evaluation layer} for
studying failure modes and model robustness before
human-subject deployment.

Cogniscope releases two complementary synthetic datasets.
\textbf{Simulation-level ($N$=200, $T$=200):} longitudinal
behavioral traces for 200 users over 200 days including daily
behavioral coherence scores, semantic drift, and nine behavioral
engagement metrics per video. Latent simulated risk state labels
(Healthy/MCI/EarlyAD/ModAD/SevAD) are retained only for
evaluation. Total: $200 \times 200 \times 5 = 200{,}000$
interaction records. \textbf{Deployment-level ($N$=504
sessions):} simulated sessions across nine behavioral profiles
($\approx$56 per profile), each containing $\approx$10
question-response records (5,040 total). Fields mirror the
Chrome extension schema: \texttt{learner\_id},
\texttt{session\_id}, \texttt{video\_topic},
\texttt{question\_type}, \texttt{question\_difficulty},
\texttt{delay\_condition}, \texttt{answer\_correct},
\texttt{response\_time\_seconds},
\texttt{video\_completion\_rate}, \texttt{pause\_count},
\texttt{replay\_count}, \texttt{skip\_count},
\texttt{missed\_question}, \texttt{attention\_noise\_level}.
Profile definitions and expected learner-status labels are in
Appendix~\ref{app:taxonomy}.

\paragraph{Artifact summary.}
Table~\ref{tab:artifacts} summarizes all released components.

\begin{table}[h]
\centering\small
\setlength{\tabcolsep}{3pt}
\caption{Cogniscope released artifacts.}
\label{tab:artifacts}
\begin{tabular}{p{2.8cm}cp{1.8cm}p{4.2cm}}
\toprule
\textbf{Artifact} & \textbf{Released} &
\textbf{Format} & \textbf{Purpose} \\
\midrule
Simulation engine   & Yes & Python &
  Generate longitudinal traces under configurable
  progression and noise \\
Simulation dataset  & Yes & CSV &
  200 users $\times$ 200 days $\times$ 5 videos;
  200,000 records \\
Deployment dataset  & Yes & CSV &
  504 schema-aligned sessions; 9 behavioral profiles \\
Chrome extension    & Yes & Manifest V3 &
  Browser-based YouTube telemetry and MCQ collection \\
Relational schema   & Yes & PostgreSQL DDL &
  18-table longitudinal monitoring schema \\
Evaluation scripts  & Yes & Python &
  ERDE, TTD, fixed-FPR, baseline models \\
Documentation       & Yes & README / metadata &
  Installation, dataset card, usage, data format,
  ethical constraints \\
\bottomrule
\end{tabular}
\end{table}

\section{Results}
\label{sec:results}

\textbf{Reporting note.} Simulation-level results are mean
$\pm$ std over 5 independent runs (different random seeds).
Variability sources: user trajectory sampling, noise injection,
and train/validation splits.

\subsection{Simulation-Level Results}

\paragraph{Coherence separability under drift and noise.}
Table~\ref{tab:coherence_noise} reports behavioral coherence
scores under clean and noisy conditions across simulated risk
states. Coherence decreases monotonically with progression
severity: Healthy users maintain high coherence ($0.880 \pm
0.001$), while EarlyAD users show substantially lower scores
($0.486 \pm 0.001$), confirming that the behavioral coherence
formula captures state-dependent engagement patterns under the
simulation priors. MCI users occupy an intermediate range
($0.692 \pm 0.001$), consistent with the heterogeneous and
transitional nature of the prodromal simulated state. Noise
injection produces minimal degradation, reflecting the
robustness of the weighted coherence formula to small
perturbations. Appendix~\ref{app:linguistic} reports the
component-level accuracy, consistency, and coherence values
underlying these aggregate trends.

\begin{table}[h]
\centering\small
\caption{Behavioral coherence under clean vs.\ noisy
  conditions (mean $\pm$ std, 5 runs). Labels denote
  simulated risk states, not clinical diagnoses.}
\label{tab:coherence_noise}
\begin{tabular}{lccc}
\toprule
\textbf{Simulated State} & \textbf{Clean} & \textbf{Noisy} &
\textbf{Drop (\%)} \\
\midrule
Healthy & $0.880 \pm 0.001$ & $0.850 \pm 0.001$ &
  $3.5 \pm 0.1$ \\
MCI     & $0.692 \pm 0.001$ & $0.685 \pm 0.001$ &
  $1.1 \pm 0.1$ \\
EarlyAD & $0.486 \pm 0.001$ & $0.488 \pm 0.001$ &
  $-0.4 \pm 0.1$ \\
\bottomrule
\end{tabular}
\end{table}

Separability analysis using Cohen's $d$
(Table~\ref{tab:drift_stats}) confirms that coherence provides
strong separation between all adjacent simulated state pairs.
Behavioral entropy also discriminates states effectively, while
drift rate mirrors coherence separability, as expected given
that drift is defined as $1 - C_{u,d}$.

\begin{table}[h]
\centering\small
\caption{Simulated state separability under drift and noise.
  All $p$-values $< 0.001$. Labels denote simulated risk
  states.}
\label{tab:drift_stats}
\begin{tabular}{llc}
\toprule
\textbf{Feature} & \textbf{Comparison} &
\textbf{Cohen's $d$} \\
\midrule
Coherence Mean     & Healthy vs MCI & 2.76 \\
                   & MCI vs EarlyAD & 2.58 \\
Behavioral Entropy & Healthy vs MCI & 1.76 \\
                   & MCI vs EarlyAD & 0.95 \\
Slope (Drift Rate) & Healthy vs MCI & 2.76 \\
                   & MCI vs EarlyAD & 2.58 \\
\bottomrule
\end{tabular}
\end{table}

\begin{figure}[h]
  \centering
  \includegraphics[width=0.68\columnwidth]{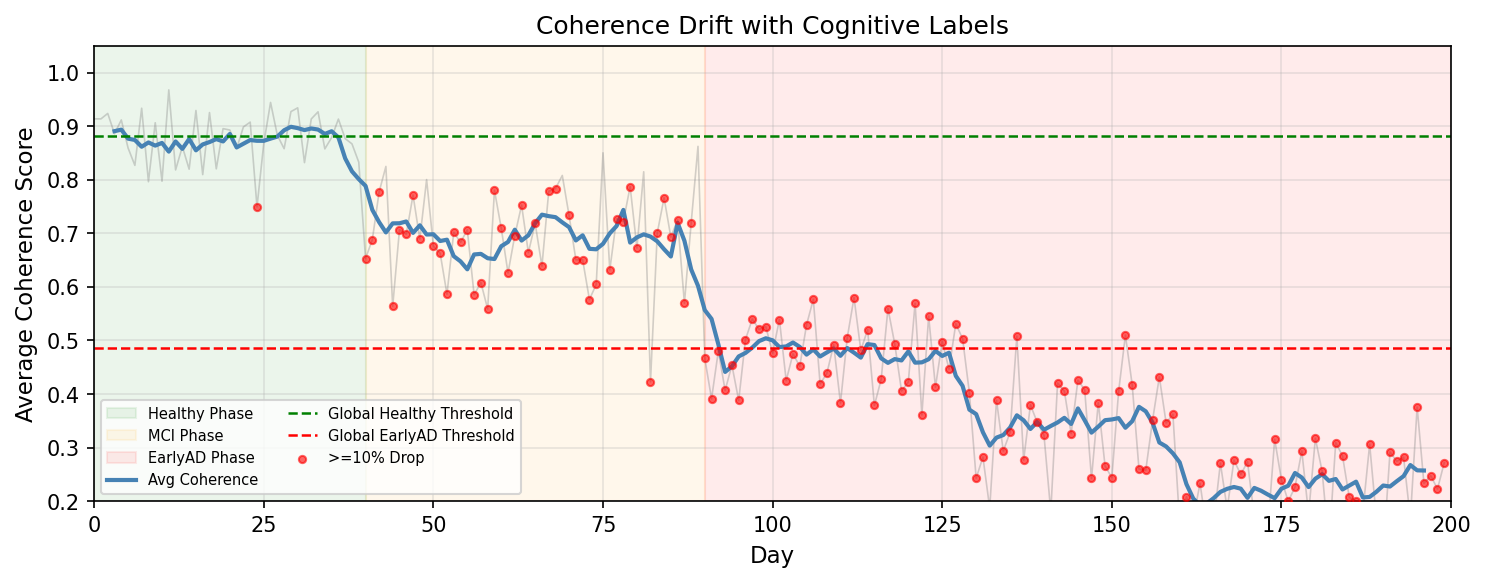}
  \caption{Coherence trajectory for a representative simulated
    user: Healthy $\rightarrow$ MCI $\rightarrow$ EarlyAD.
    Labels denote simulated risk states. Dashed lines indicate
    state-specific global thresholds.}
  \label{fig:drift_example1}
\end{figure}

\paragraph{Longitudinal coherence trajectories.}
Over 200 simulated days, Healthy users maintain stable
coherence near the global healthy threshold, MCI users show
gradual decline with heterogeneous trajectories, and EarlyAD
users exhibit sustained deterioration below the EarlyAD
threshold (Figure~\ref{fig:drift_example1}). This pattern
motivates continuous monitoring over episodic screening and
validates that the simulation produces internally consistent
progression dynamics.

\paragraph{Ablation study.}
Table~\ref{tab:ablation} reports classification performance
under four conditions. The ceiling performance of the full and
behavior-only settings should \emph{not} be interpreted as
evidence of clinical predictive validity. Rather, it confirms
that the simulator produces internally consistent trajectories
under the default priors, a necessary sanity check for any
synthetic benchmark. The more informative result is the
coherence-only setting, which removes many direct behavioral
indicators and tests whether an aggregated behavioral-coherence
signal retains useful temporal information. We therefore treat
full-model results as a \emph{simulator sanity check} and
coherence-only, noisy, and challenge-split evaluations as the
primary benchmark conditions.

\begin{table}[h]
\centering\small
\caption{Ablation study (mean $\pm$ std, 5 runs, 70/30 split)
  and deployment classifier performance. Full model and
  behavior-only achieve ceiling performance confirming
  simulator internal consistency; coherence-only results
  reflect realistic partial-feature evaluation.}
\label{tab:ablation}
\begin{tabular}{lccc}
\toprule
\textbf{Setting / Metric} & \textbf{Accuracy} &
\textbf{F1 (MCI)} & \textbf{F1 (EarlyAD)} \\
\midrule
\multicolumn{4}{l}{\textit{Ablation Study}} \\
Full Model (sanity check) & $1.000 \pm 0.000$ &
  $1.000 \pm 0.000$ & $1.000 \pm 0.000$ \\
Coherence Only (primary) & $0.886 \pm 0.002$ &
  $0.819 \pm 0.003$ & $0.898 \pm 0.003$ \\
Behavior Only (sanity check) & $1.000 \pm 0.000$ &
  $1.000 \pm 0.000$ & $1.000 \pm 0.000$ \\
No Noise Injection & $1.000 \pm 0.000$ &
  $1.000 \pm 0.000$ & $1.000 \pm 0.000$ \\
\midrule
\multicolumn{4}{l}{\textit{Deployment Classifier}} \\
Macro $F_1$        & \multicolumn{3}{c}{0.466} \\
Macro Precision    & \multicolumn{3}{c}{0.702} \\
Macro Recall       & \multicolumn{3}{c}{0.484} \\
Cohen's $\kappa$   & \multicolumn{3}{c}{0.42} \\
\bottomrule
\end{tabular}
\end{table}

\begin{figure}[h]
  \centering
  \includegraphics[width=0.66\linewidth]{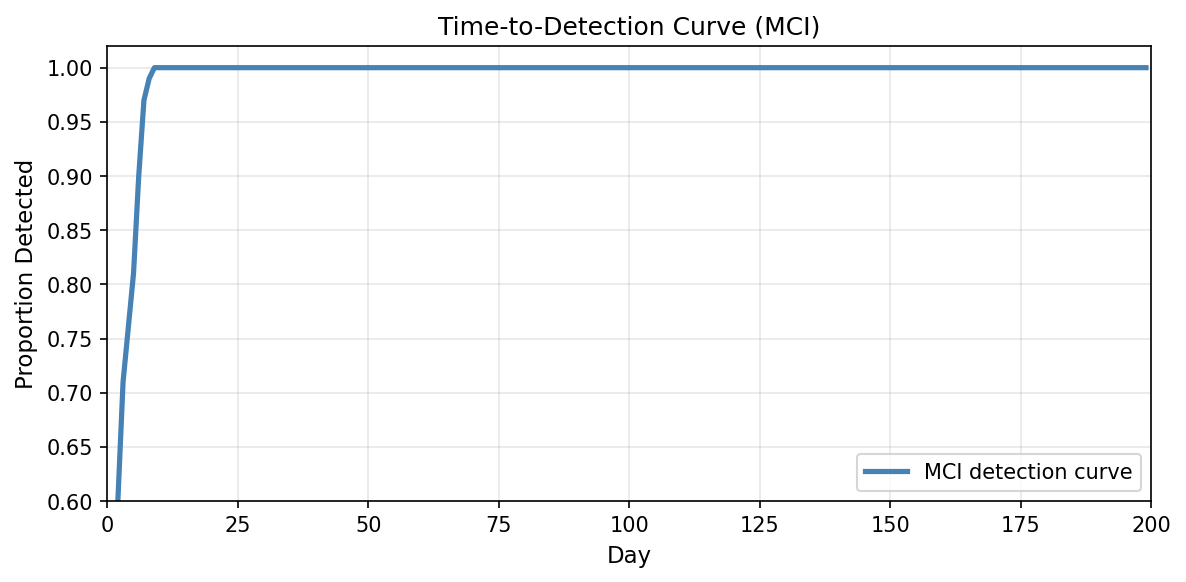}
  \caption{Time-to-detection curves under fixed coherence
    threshold within the simulation environment.}
  \label{fig:ttd}
\end{figure}

\paragraph{Time-to-detection.}
The behavioral coherence threshold detector identifies
simulated risk-state onset rapidly: over 95\% of simulated
users are detected within 10 days of MCI-state onset under a
fixed coherence threshold of 0.65 (Figure~\ref{fig:ttd}).
This rapid detection reflects the simulation's well-separated
state distributions and serves as an upper-bound benchmark for
real-world systems operating under greater uncertainty.

\subsection{Deployment-Level Results}

The rule-based learner-status classifier achieves macro $F_1 =
0.47$ (Precision $= 0.70$, Recall $= 0.48$, Cohen's $\kappa =
0.42$) on the 504-session synthetic dataset. The moderate
recall reflects boundary ambiguity between adjacent
learner-status categories, which share overlapping feature
ranges by design. Full per-class results are in
Appendix~\ref{app:classifier}. These results highlight a key
finding: rule-based classifiers with hard thresholds are
insufficient for naturalistic behavioral data, motivating the
use of learned temporal models as the primary evaluation
target.

\section{Responsible Design}
\label{sec:responsible}

Cogniscope is designed as a \emph{human-review support analytics
framework}. The \texttt{frontend\_summary\_view} exposes only
neutral behavioral statistics (``stable,'' ``minor change
observed,'' ``pattern worth monitoring''); disease names, raw
risk scores, and diagnostic conclusions are never surfaced to
end users. Prediction scores and reviewer flags are stored with
\texttt{visible\_to\_clinician\_only = TRUE} and restricted via
Supabase row-level security. Elevated risk signals trigger
documented safety-action workflows. Two population-specific
pathways are supported: an \emph{older-adult cognitive pathway}
emphasizing delayed recall, response latency trends, and replay
behavior; and a \emph{youth mental-health pathway} emphasizing
circadian viewing patterns, content sentiment trends, and
mood-snapshot consistency.

\section{Limitations}
\label{sec:limitations}

The principal limitation is that the current benchmark
evaluates model behavior under controlled synthetic assumptions
rather than real-world clinical validity; therefore, high
performance on Cogniscope should be interpreted as evidence of
robustness to specified simulation conditions, not as evidence
of diagnostic accuracy.
Both datasets are fully synthetic; the simulation uses a
monotonic progression model that does not capture remission or
fluctuating trajectories, and neither dataset captures
demographic variation. The behavioral coherence formula
(Eq.~\ref{eq:coherence}) is a proxy measure; future work will
integrate transcript-based embeddings when textual summaries
are available from the Chrome extension. The ceiling
performance of the full model on synthetic data reflects the
deterministic structure of simulation priors rather than true
generalization; real-world validation is required to assess
performance under naturalistic noise and heterogeneity.
Behavioral distributions (Table~\ref{tab:behaviors}) are
illustrative and calibrated against public engagement
statistics but cannot fully capture platform-specific
dynamics~\citep{pew,Milne2022BigDataSoc}. Quantitative
validation against real longitudinal cohorts (eRisk, ADReSS)
is a priority for future
work~\citep{LosadaCrestani2016,luz2020adress,Lanzi2023}. The
simulation cohort ($N=200$, $T=200$) and deployment dataset
($N=504$) are modest; future work will scale both and integrate
neural sequence models. The extension currently targets
YouTube; generalization requires additional content script
development. LLM-generated questions may contain factual errors
despite validation. Cogniscope does not diagnose cognitive
impairment; all outputs are screening signals intended to
support, not replace, clinical judgment.

\section{Conclusion}
\label{sec:conclusion}

We presented Cogniscope, a synthetic longitudinal benchmark and
browser-based evaluation framework for early-risk cognitive AI
systems. The simulation level provides a configurable,
privacy-preserving engine for controlled experimentation on
temporal modeling, multimodal signal integration, and
robustness under realistic uncertainty. The deployment level
provides a Chrome extension implementation, an 18-table
relational schema, and a 504-session synthetic benchmark with a
transparent rule-based classifier. The behavioral coherence
formula achieves $F_1(\text{MCI}) = 0.819 \pm 0.003$ and
$F_1(\text{EarlyAD}) = 0.898 \pm 0.003$ under coherence-only
evaluation on synthetic data, demonstrating that response-level
behavioral signals carry discriminative information under
controlled simulation conditions. Together, these levels address
a critical gap: no existing resource simultaneously provides
all four components of a complete synthetic evaluation
ecosystem. By releasing all artifacts openly, Cogniscope enables
the research community to develop, evaluate, and compare
cognitive assessment models in ecologically motivated settings
while maintaining a strict human-review posture and clear
separation from clinical claims. Future work will validate the
framework against real user data collected via the Chrome
extension and extend the simulation to non-monotonic progression
trajectories~\citep{Mueller2005ADNI,Topol2019NatMed}.

\section*{Ethical Considerations and Broader Impact}

The simulation operates exclusively on synthetic data; no real
user data is collected. For the deployed extension, all user
identifiers are anonymized, content payloads are stripped
locally before transmission, sensitive domains are excluded via
a hardcoded blacklist, data at rest is encrypted (AES-256),
and data in transit uses TLS 1.3. Extension users are informed
of data collection at onboarding; participation is voluntary.
Predictions must not be interpreted as clinical diagnoses or
used for high-stakes decision-making without validated medical
evidence~\citep{blasi2005assessment,petersen2014lancet}.
Passive behavioral monitoring carries over-surveillance risks;
Cogniscope mitigates this through restricted prediction access,
neutral frontend language, and required human review of any
clinical interpretation. Future real-world data collection must
address age, ethnicity, language, and device diversity, and must
comply with IRB requirements, GDPR, and HIPAA~\citep{
Milne2022BigDataSoc,Smedinga2018JAD}.

\section*{Acknowledgements}
A large language model assistant was used to assist in drafting
and editing text and debugging code. Grammar-checking tools
were used for proofreading. All outputs were reviewed and
edited by the authors; all scientific claims and citations
remain the authors' sole responsibility. All citations were
manually verified against primary sources.

\bibliographystyle{plainnat}
\bibliography{references}

\appendix

\section{Algorithms}
\label{app:algorithms}

This appendix presents the two core algorithms used in
Cogniscope: the daily user simulation procedure
(Algorithm~\ref{alg:simulation}) and the noisy training
protocol (Algorithm~\ref{alg:training}).

\begin{algorithm}[h]
\caption{Daily User Simulation in Cogniscope}
\label{alg:simulation}
\begin{algorithmic}[1]
\REQUIRE User $u$, Day $d$, Progression profile $P_u$,
  Mode $\in \{\text{full}, \text{no-label}\}$
\STATE Initialize: $\mathit{useLLM} \leftarrow
  (u \in \mathcal{U}_\text{LLM})$
\STATE Determine latent state $L_d \leftarrow f(P_u, d)$
\STATE Sample $k$ video titles $V_d = \{v_1, \ldots, v_k\}$
\FOR{each video $v \in V_d$}
  \IF{$\mathit{useLLM}$}
    \STATE Query LLM (full: parameterized prompt; no-label:
      neutral prompt); fallback to template on API failure
  \ELSE
    \STATE $s \leftarrow$ template summary with state-dependent
      noise (full) or neutral template (no-label)
  \ENDIF
  \STATE Sample $\{\mathrm{acc}, t, \mathrm{skip},
    \mathrm{cons}\} \sim \mathrm{Priors}(L_d)$
  \STATE $C_{u,d,v} \leftarrow w_1 \cdot \mathrm{acc}
    + w_2 \cdot e^{-t/60}
    + w_3 \cdot (1-\mathrm{skip})
    + w_4 \cdot \mathrm{cons}$
  \STATE Sample engagement $\{\mathrm{watch, pause, replay,
    like, share}\} \sim \mathrm{Priors}(L_d)$
  \STATE Append $\{u, d, v, C_{u,d,v}, \mathrm{behaviors}\}$
    to session log
\ENDFOR
\STATE Store $L_d$ separately for evaluation only
\end{algorithmic}
\end{algorithm}

\begin{algorithm}[h]
\caption{Training with Noisy and Confounded Data}
\label{alg:training}
\begin{algorithmic}[1]
\REQUIRE Clean dataset $\mathcal{D}$ with user-level labels
\FOR{each feature vector $f_{u,d}$}
  \STATE $f_{u,d} \leftarrow f_{u,d} + \mathcal{N}(0,\sigma^2)$,
    $\sigma \in \{0.05, 0.1, 0.2, 0.3\}$
  \STATE Flip binary outcomes with $p \sim \mathcal{U}(0,0.1)$
\ENDFOR
\STATE Apply user-level confounds; clip to realistic bounds
\STATE Split 70/30 (no leakage); train; evaluate (Acc, F1, ERDE)
\ENSURE $\mathcal{D}'$, trained model, robustness metrics
\end{algorithmic}
\end{algorithm}

\section{Hyperparameter Details}
\label{app:hyperparams}

All temporal models trained with Adam ($\text{lr}=10^{-3}$),
batch size 32, early stopping (patience=10). GRU: hidden 64,
2 layers. TCN: kernel 3, 64 filters, 4 layers. Transformer:
2 heads, 64-dim, 2 encoder layers. LR: $\ell_2$, $C=1.0$.
Hardware: single CPU (Intel Core i7, 16GB RAM); total runtime
$\approx$4 hours for $N=200$, $T=200$.

\section{LLM Usage in Data Generation}
\label{app:llm}

Summaries for $|\mathcal{U}_\text{LLM}|=100$ users were
generated via Groq API (Llama-3-based). No-label mode prompts
contain only video title and category; no cognitive state is
exposed. Remaining 100 users use template-based generation
with state-dependent noise, reducing dependence on any single
LLM architecture.

\section{Full Schema Table Inventory}
\label{app:schema_full}

\begin{table}[h]
\centering\small
\caption{All 18 Cogniscope schema tables.}
\label{tab:schema_summary}
\begin{tabular}{lll}
\toprule
\textbf{Table} & \textbf{Layer} & \textbf{Primary Purpose} \\
\midrule
\texttt{user\_info} & User & Root user profile \\
\texttt{population\_profile} & User & Monitoring pathway \\
\texttt{video\_info} & Session & Per-session watch record \\
\texttt{video\_content\_metadata} & Session &
  Content properties \\
\texttt{video\_interaction\_events} & Interaction &
  Raw telemetry \\
\texttt{user\_response} & Interaction &
  MCQ answers \& latency \\
\texttt{question\_presentation\_log} & Interaction &
  Display context \\
\texttt{question\_bank} & Assessment &
  Annotated questions \\
\texttt{assessment\_schedule} & Assessment &
  Baseline/follow-up \\
\texttt{assessment\_result} & Assessment &
  Longitudinal scores \\
\texttt{derived\_feature\_store} & Modeling &
  Engineered features \\
\texttt{longitudinal\_summary} & Modeling &
  Rolling change metrics \\
\texttt{clinical\_label} & Modeling &
  Ground-truth labels \\
\texttt{label\_evidence\_link} & Modeling &
  Label traceability \\
\texttt{risk\_signal} & Safety & Risk indicators \\
\texttt{safety\_action\_log} & Safety &
  Follow-up records \\
\texttt{prediction\_result} & Safety &
  Restricted model outputs \\
\texttt{frontend\_summary\_view} & Safety &
  Neutral user statistics \\
\bottomrule
\end{tabular}
\end{table}

\section{Schema DDL Excerpt}
\label{app:ddl}

\begin{verbatim}
CREATE TABLE video_info (
  session_id        UUID PRIMARY KEY
                    DEFAULT gen_random_uuid(),
  user_id           UUID NOT NULL
                    REFERENCES user_info(user_id),
  video_id          TEXT NOT NULL,
  video_title       TEXT,
  video_duration    INTEGER,
  timestamp_started TIMESTAMPTZ,
  timestamp_stopped TIMESTAMPTZ,
  completion_rate   FLOAT,
  device_type       TEXT,
  session_mood_pre  TEXT,
  session_mood_post TEXT
);
CREATE TABLE video_interaction_events (
  event_id           UUID PRIMARY KEY
                     DEFAULT gen_random_uuid(),
  session_id         UUID NOT NULL
                     REFERENCES video_info(session_id),
  user_id            UUID NOT NULL
                     REFERENCES user_info(user_id),
  event_type         TEXT NOT NULL,
  event_ts           TIMESTAMPTZ NOT NULL,
  video_time_seconds FLOAT,
  delta_seconds      FLOAT,
  cognitive_context_tag TEXT,
  idle_time_before   FLOAT
);
CREATE TABLE derived_feature_store (
  feature_record_id    UUID PRIMARY KEY
                       DEFAULT gen_random_uuid(),
  user_id              UUID NOT NULL
                       REFERENCES user_info(user_id),
  session_id           UUID
                       REFERENCES video_info(session_id),
  feature_window_start TIMESTAMPTZ,
  feature_window_end   TIMESTAMPTZ,
  population_group     TEXT,
  feature_name         TEXT NOT NULL,
  feature_value        FLOAT NOT NULL
);
\end{verbatim}

\section{Learner-Status Taxonomy}
\label{app:taxonomy}

\begin{table}[h]
\centering\small
\caption{Nine learner-status categories (priority order).
  \textbf{Not clinical diagnoses.}}
\label{tab:learner_status}
\begin{tabular}{p{2.6cm}p{5.0cm}p{3.0cm}}
\toprule
\textbf{Status} & \textbf{Rule (abbreviated)} &
\textbf{System Action} \\
\midrule
Low-Engagement & completion $<$ 0.60 OR missed\_q $\geq$ 0.40
  OR avg\_skip $\geq$ 3 & Shorter content \\
Fast but Inaccurate & median\_rt $\leq$ 6s AND acc $<$ 0.60 &
  Slow-down prompt \\
Delayed Recall Weakness & imm\_acc $\geq$ 0.75 AND
  del\_acc $<$ 0.60 AND drop $\geq$ 0.25 &
  Spaced repetition \\
High Cognitive Load & acc $<$ 0.60 AND rt $\geq$ 20s AND
  completion $\geq$ 0.70 & Simpler content \\
Attention-Fluctuating & acc\_var $\geq$ 0.25 OR
  rt\_var $\geq$ 10 & Shorter segments \\
Strong Retention & acc $\geq$ 0.80 AND del\_acc $\geq$ 0.75
  AND drop $\leq$ 0.15 & No intervention \\
Stable Learner & acc $\geq$ 0.65 AND del\_acc $\geq$ 0.60 &
  Maintain difficulty \\
Slow but Accurate & rt $\geq$ 20s AND acc $\geq$ 0.70 &
  Extended time \\
Needs Review & acc $<$ 0.65 OR del\_acc $<$ 0.60 &
  Review segments \\
\bottomrule
\end{tabular}
\end{table}

\section{Deployment-Level Classifier Results}
\label{app:classifier}

\begin{table}[h]
\centering\small
\caption{Rule-based classifier on 504-session synthetic dataset
  (9 classes). Macro $F_1 = 0.47$, Precision $= 0.70$,
  Recall $= 0.48$, Cohen's $\kappa = 0.42$.}
\label{tab:classifier_results}
\begin{tabular}{lrrr}
\toprule
\textbf{Learner Status} & \textbf{P} & \textbf{R} &
\textbf{$F_1$} \\
\midrule
Attention-Fluctuating    & 1.00 & 0.11 & 0.19 \\
Delayed Recall Weakness  & 0.76 & 0.55 & 0.64 \\
Fast but Inaccurate      & 1.00 & 0.86 & 0.92 \\
High Cognitive Load      & 0.89 & 0.73 & 0.80 \\
Low-Engagement           & 0.40 & 0.89 & 0.56 \\
Needs Review             & 0.30 & 0.57 & 0.39 \\
Slow but Accurate        & 1.00 & 0.02 & 0.04 \\
Stable Learner           & 0.17 & 0.34 & 0.23 \\
Strong Retention         & 0.80 & 0.29 & 0.42 \\
\midrule
\textbf{Macro Avg} & \textbf{0.70} & \textbf{0.48} &
  \textbf{0.47} \\
\bottomrule
\end{tabular}
\end{table}

\section{Linguistic Degradation Metrics}
\label{app:linguistic}

\begin{table}[h]
\centering\small
\caption{Coherence component metrics relative to baseline
  (days 1--5), mean $\pm$ std over 5 runs. Labels denote
  simulated risk states.}
\label{tab:linguistic_metrics}
\begin{tabular}{lccc}
\toprule
\textbf{Simulated State} & \textbf{Accuracy} &
\textbf{Consistency} & \textbf{Coherence Score} \\
\midrule
Healthy & $0.850 \pm 0.008$ & $0.880 \pm 0.006$ &
  $0.880 \pm 0.001$ \\
MCI     & $0.650 \pm 0.012$ & $0.620 \pm 0.012$ &
  $0.692 \pm 0.001$ \\
EarlyAD & $0.420 \pm 0.014$ & $0.380 \pm 0.014$ &
  $0.486 \pm 0.001$ \\
\bottomrule
\end{tabular}
\end{table}

\section{Prior Work Comparison}
\label{app:prior_work}

\begin{table}[h]
\centering\small
\caption{Representative prior work and Cogniscope's relation.}
\label{tab:prior_work}
\begin{tabular}{p{2.2cm}p{2.2cm}p{5.8cm}}
\toprule
\textbf{Study} & \textbf{Modality} & \textbf{Key Finding /
  Relation} \\
\midrule
\citet{Fraser2015JAD}
  & Narrative speech & $F_1 \sim 0.72$ AD vs.\ HC; Cogniscope
    adds engagement + temporal modeling \\
\citet{Balagopalan2021Frontiers}
  & Acoustic + text & $F_1 \sim 0.74$; Cogniscope extends to
    passive video interaction \\
\citet{Seelye2015DADM}
  & Mouse tracking & Motor patterns indicate MCI; Cogniscope
    captures analogous video telemetry \\
\citet{pahar2025cognospeak}
  & Conv.\ speech & $F_1 = 0.873$; requires dedicated sessions
    vs.\ Cogniscope's passive capture \\
\textbf{Cogniscope (ours)}
  & Synthetic video telemetry + behavioral coherence &
    Synthetic longitudinal benchmark; 18-table schema;
    200K + 5K records; coherence-only
    $F_1(\text{MCI})=0.819$ on simulated data \\
\bottomrule
\end{tabular}
\end{table}

\end{document}